# Weak optical modes for high-density and low-loss photonic circuits

Hamed Nikbakht[1], Bob van Someren[3], Manfred Hammer[2] and B. Imran Akca[1,*]

[1]*LaserLaB, Department of Physics and Astronomy, VU University Amsterdam, De Boelelaan 1081, 1081 HV Amsterdam, The Netherlands*
[2] *Elf Software, Mullerkade 667, 3024 EP, Rotterdam, The Netherlands*
[3]*Paderborn University, Theoretical Electrical Engineering, Warburger Straße 100, 33098 Paderborn, Germany*
*\*b.i.avci@vu.nl*

**Abstract:** Dielectric optical waveguides constitute the main building blocks of photonic integrated circuits (PICs). Channels with high refractive index contrast can provide very compact PIC components whereas structures with lower index exhibit less propagation loss. A hybrid concept that can combine the best of high- and low-index materials is highly required. Here, we devise a new approach to realize compact and low-loss hybrid optical waveguides based on the interaction of weak optical modes. This is a rather universal approach that can be applied to a wide range of optical materials. To prove the principle, the hybrid waveguide structure is formed by combining a low-index polymer and a thin layer of silicon nitride. For this material combination, a minimum bending radius of 90 µm (for a bending loss of 0.005 dB/90º) and an upper limit for the optical loss of 0.7 dB/cm are achieved. The viability of this platform is demonstrated through a series of high-performance novel PIC components. This hybrid waveguide platform enabled by a powerful and simple design concept holds great promise for high-density and low-loss PICs.

## 1. Introduction

Photonic integrated circuits (PICs) are key elements in various fields of current and future technological applications. Among these are quantum computers [1], neuromorphic networks [2], portable imaging systems [3], and disposable sensors [4]. Many PIC devices rely on dielectric optical waveguides as their main building block. As a prerequisite for high-density photonic integration, channels with strong confinement enable small cross-section dimensions and small bending radii of curved waveguides. For strong confinement, high-index materials, such as silicon and indium phosphate, are preferred; however, this leads to higher optical losses due to scattering caused by surface roughness and refractive index steps. Alternatively, low-index-contrast optical waveguides can provide much lower losses, but usually at the expense of large device sizes. Therefore new concepts of optical waveguides are needed that can combine the best of high- and low-index contrast materials. One of the earliest attempts toward this goal was made in 1974 by placing a low-index photonic material (e.g. silicon dioxide) on top of a high-index guiding layer to create lateral optical confinement [5]. Despite its great potential, this concept (i.e. strip-loaded optical waveguides) has remained less favorable due to large optical mode sizes and thereby large device footprints. Moreover, the availability of loading materials with different refractive index values has limited the design flexibility. Recently, this concept regained attention through the work of Zou *et al.* [6]. They proposed a strip-loaded waveguide geometry based on the concept of bound states in the continuum (BIC), which was first proposed by von Neumann and Wigner in 1929 [7]. Based on the BIC concept, Yu *et al.* demonstrated a hybrid waveguide by combining a low-index polymer with lithium niobate on an insulator [8]. In this concept, the waveguide width was varied to find the bound states; however, the effect of the guiding layer thickness was not properly addressed. Moreover, the nature of the BIC concept allows minimum loss at only certain bending radius values, which limits its use in applications that rely critically on a specific free spectral range of some optical cavities.

Here, we exploit the potential of strip-loaded optical waveguides through a new approach that makes it possible to realize compact and low-loss hybrid optical waveguides [9]. In essence, our approach couples two vertically weakly guiding regions, with near-cutoff optical modes, into an optical waveguide with strong mode confinement. Figure 1 illustrates the concept. We apply our approach to the fabrication of single-mode hybrid waveguides for operation in the telecom wavelength range, on the basis of a low-index polymer and a thin layer of silicon nitride ($Si_3N_4$). Merely single-step UV lithography and a standard etchant are required, which shortens fabrication time to less than an hour. In contrast to existing strip-loaded waveguides [10-13], we achieve a significantly lower optical loss and a very compact bend radius. To demonstrate the feasibility of this platform, we realized a series of high-performance PIC devices including an elliptical ring resonator, an arrayed waveguide grating (AWG), a new type of ultra-broadband optical splitter, and two new types of coupled microcavity resonators.

## 2. Methodology

### 2.1 Hybrid waveguide concept

Realizing compact PICs in low-index contrast systems can be challenging. We address that task by means of suitably customized hybrid waveguides, as illustrated in Fig. 1(a). The hybrid waveguide is formed by combining a high-index guiding layer of refractive index $n_g$ with a lower-index loading layer of refractive index $n_p$. The thicknesses of the guiding layer, $t_g$, and of the loading layer, $t_p$, have to be chosen carefully, with an aim of effecting a suitable distribution of the mode field among the layers, that leads to configurations with reduced contributions of material and radiation losses to the overall modal attenuation.

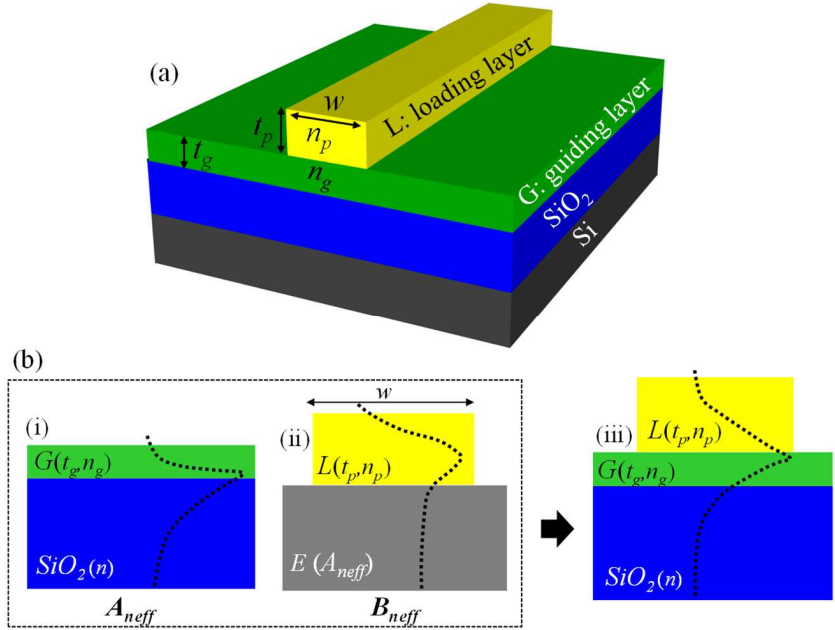

**Figure 1. (a)** Schematic of the hybrid optical waveguide. **(b)** The weakly-confined vertical mode profiles in the (i) loading and (ii) guiding layers of the hybrid waveguide are shown with dotted black lines. The right image (iii) shows the confined mode formed by combining these weakly-confined vertical optical modes.

In contrast to traditional design approaches, where the waveguide geometry is analyzed as a whole, we divide the optical waveguide into two sub-systems (Fig. 1(b)). Guiding and loading layers are treated as individual guiding systems with a certain coupling between them. Since the refractive indices of these layers are relatively different from each other, the interaction between them is weak. To increase the coupling, a strong mutual overlap of the light (vertically) between the two systems is needed. This can be achieved by choosing the layer thicknesses thin enough (i.e. just above cut-off) such that light is only weakly confined to the individual layers whereas it is highly confined in the combined layer stack. In essence, we strongly couple weak modes of two distinct waveguides vertically by evanescent fields [14]. Assuming that material absorption is mainly present in the medium of the loading layer, this weak confinement approach automatically reduces the modal attenuation.

For further design, we specialize in a target vacuum wavelength of 1550 nm, and restrict things to TE polarization. We first consider the guiding layer only with a refractive index of $n_g$ and thickness of $t_g$ (Fig. 1(b), i), sandwiched between a substrate medium (SiO$_2$) with a refractive index of 1.467 below, and the air above. Figure 2(a) shows the effective refractive index $A_{neff}$ of the fundamental TE mode of the respective slab waveguide as a function of $n_g$ and $t_g$. The white-dotted line indicates weakly guiding, close-to-cutoff configurations for further use in our waveguides.

The same procedure is applied to the loading layer (Fig. 1(b), ii) to identify suitable thicknesses $t_p$ for different values of the refractive index $n_p$. As an imitation of the final channel configuration, we here assume that the loading layer is placed on top of an effective layer $E$ with an (effective) refractive index of $A_{neff}$, where $A_{neff}$ is the effective index selected for the guiding layer in the previous step. Next, we choose the width $w$ of the loading layer in such a way that the strip waveguide of an intermediate thickness, with intermediate core refractive index, remains single mode. Figure 2(b) then shows a map of effective index values for the fundamental TE mode supported by the strip waveguide with core refractive index $n_p$ and thickness $t_p$. For the calculations, we assume a width $w$ of 1.5 µm, and the value $A_{neff} = 1.464$ associated with the formerly weakly guided configurations on the white line in Figure 2(a). The range from 1.55 to 1.7 represents typical refractive indices of commercially-available polymers. Also in Fig. 2(b), the white-dotted line identifies close-to-cutoff structures with weak vertical field confinement.

As a final step, shown in Fig. 1(b), iii, these two thin optical layers are combined into a strip-loaded waveguide with the parameters $n_g$, $t_g$, $n_p$, $t_p$ (and $w$) of the separate systems. We shall see that, as long as weakly guided configurations on the white-dotted lines are selected, this recipe leads to waveguides that support highly-confined optical modes that show only low bending losses. Just as for the calculations underlying Fig. 2(b), simulations based on a 3D beam propagation method (BPM, RSOFT Inc.) were used to rigorously characterize all final strip-loaded waveguides discussed here. A simulated bend model [15] was applied to determine bending losses.

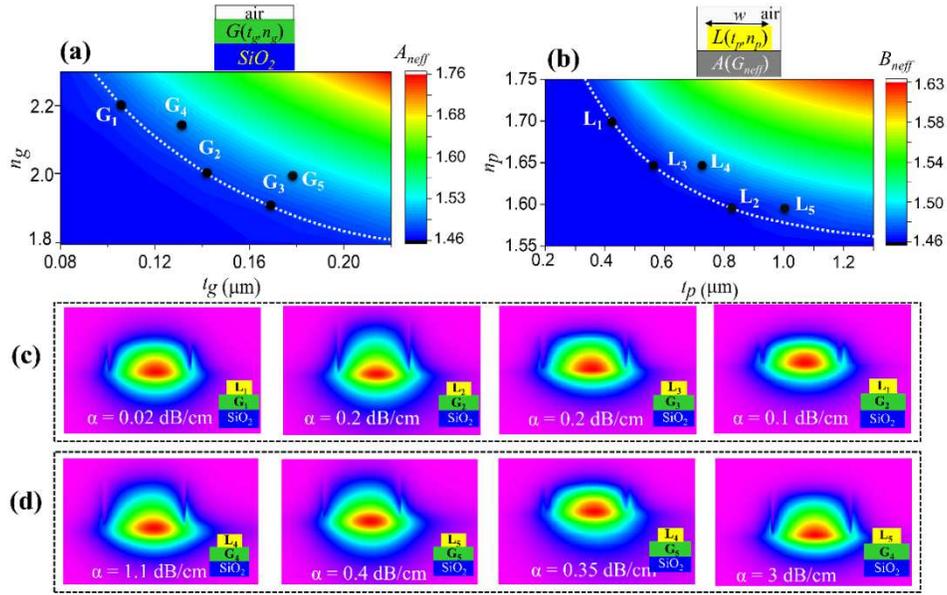

**Figure 2.** The effective refractive index values for different **(a)** $n_g$ - $t_g$ combinations and **(b)** $n_p$ - $t_p$ combinations calculated at $\lambda = 1.55$ µm for the layer stacks shown on top of each graph. The mode profiles and optical loss values (material + radiation) of the optical modes when the thickness of the guiding layer and the loading layer are **(c)** close to the cut-off thickness and **(d)** far from the cut-off thickness for the bending radius of $R$=90 µm. The $A_{neff}$ and $B_{neff}$ values are indicated with different colors. Note that an optical mode with minimum loss is obtained when the guiding and loading layer thicknesses are chosen close to the cut-off values.

We tested our approach on different material combinations. To do so, we chose five guiding (G1-G5) and five loading (L1-L5) materials and respective thicknesses as depicted with black dots in Fig. 2(a) and 2(b). We paired these materials and calculated their optical loss (bending radiation loss & absorption loss of the polymer layer, assuming a medium with an absorption coefficient of $k_{abs}$=1.25×10-6) for bend waveguides with a bending radius of 90 µm and a width $w$=1.5µm. As expected, if the $n_g$-$t_g$ and $n_p$-$t_p$ combinations are chosen from the region close to the white-dotted lines, waveguides and waveguide bends with acceptably low loss are realized. These are shown in Fig. 2(c). As negative examples, Fig. 2(d) shows several configurations chosen from outside of these regions that exhibit substantially higher losses. In short, with this new approach, we provide a quite general design strategy where our lookup table in Fig. 2(a)&(b) for various guiding and loading materials serves as a simple means for the dimensioning of low-loss channel waveguides that enable tight bends.

### 2.2 Design and fabrication

To prove the principle of our approach, we used LPCVD-deposited $Si_3N_4$ as the guiding layer and SU8 (MicroChemicals Gmbh) as the loading layer. For the guiding layer, the effective refractive index ($A_{neff}$) of the fundamental mode is calculated for the $t_g$ values ranging between 80-220 nm (Fig. 3a). As can be seen from this figure, the light mostly travels in the bottom $SiO_2$ cladding when $t_g \leq 140$ nm. So the minimum thickness that can weakly confine the light ($A_{neff}$= 1.467) in this layer is extracted as 140 nm ±5 nm. We applied the same approach to the loading layer and calculated the effective refractive index ($B_{neff}$) of the fundamental mode as given in Fig. 3b. The mode mostly travels in the bottom effective layer when $t_p \leq 1000$ nm. Therefore, the optimum value of $t_p$ is extracted as 1000 nm±20 nm, which corresponds to weak vertical confinement as well. The bending loss of the hybrid waveguide was calculated for the bending radius range of $R = 70$-110 µm (Fig. 3c). The minimum bending radius that results in a bending loss value of 0.005 dB/90⁰ was extracted as $R$ = 90 μm. The material absorption loss of 0.5 dB/cm was obtained when the extinction coefficient of $1.25×10^{-6}$ was used for the SU8 layer.

The LPCVD $Si_3N_4$ film was deposited on an 8-μm-thick thermally-oxidized silicon wafer and annealed at 1200 ºC in an $N_2$ environment for 3 hours. An air cladding was used in our current devices; however, a low-index polymer layer was also tested ($n$ = 1.1 at $λ$=1550 nm by Inkron), which worked as well. The refractive index of the thermal oxide, polymer layer, and $Si_3N_4$ layer are 1.464, 1.58, and 2.0 at $λ$=1550 nm, respectively. For the coupled cavity resonators, which operate at the 1300 nm range, the width was chosen as $w$=1.3 μm to guide only the fundamental mode. For all other devices $w$=1.5 μm. The devices were fabricated using conventional photolithographic patterning followed by developing exposed samples with 1-methoxy-2-propanol acetate. The material absorption of the SU8 layer was ~1.5 dB/cm at 1550 nm wavelength and currently, we are testing more transparent polymers to reduce the material loss (e.g. CP5, HighRI Optics or Norland optical glues). The scanning electron microscope (SEM) image of the waveguide cross-section, some of the fabricated devices, and the red light propagation in the fabricated waveguide are given in Fig. 3d, Fig. 3e&f, and Fig. 3g, respectively.

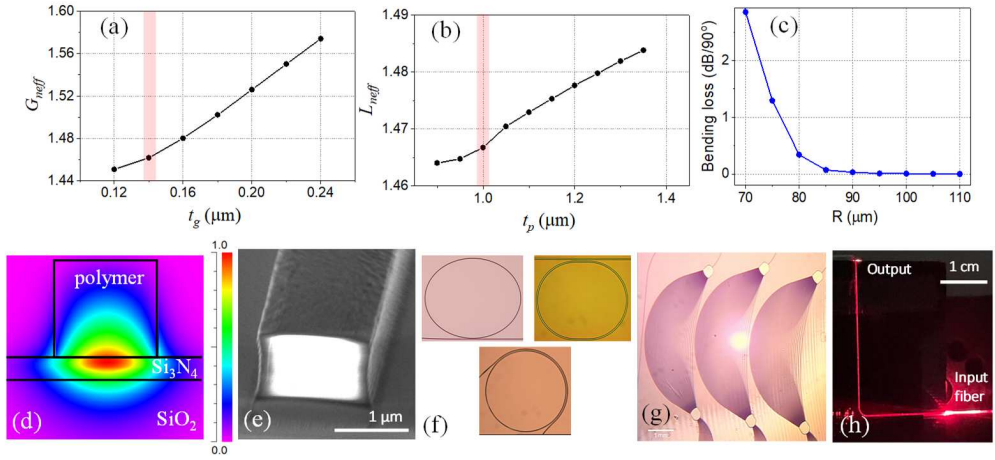

**Figure 3.** The simulated effective refractive indices of the hybrid waveguide when the loading layer is SU8 and the guiding layer is $Si_3N_4$ are shown in Fig. 1b **(a)** (i) and **(b)** (ii). The optimum values of the $Si_3N_4$ and the SU8 layers were extracted as $t_g$ = 140±5 nm and $t_p$ = 1000±50 nm, respectively. **(c)** Bending loss versus bending radius for the hybrid waveguide structure given in (iii) with $t_g$ = 140 nm and $t_p$ = 1000 nm. **(d)** The mode profile. **(e)** The SEM image of the waveguide cross-section. Optical microscope images of the fabricated **(f)** resonators and **(g)** arrayed waveguide grating. **(h)** Red light propagation in the fabricated waveguide.

*2.3 Optical loss measurements*

It is possible to infer the optical loss ($α$) of a waveguide from the spectral characteristics of the microring resonator using the formula below [16]:

$$\alpha(dB/cm) = 4.34 \times \frac{\lambda_0}{Q_{int} \times FSR \times R} \quad (1)$$

where $λ_0$ is the resonant wavelength, $FSR$ is the free spectral range, $R$ is the radius of the microring resonator and $Q_{int}$ is the intrinsic quality factor, which is defined as [17]:

$$Q_{int} = \frac{2}{(1+\sqrt{T_0})} \times \frac{\lambda_0}{FWHM} \qquad (2)$$

Here $T_0$ is the transmitted optical power at the resonant wavelength $\lambda_0$ and *FWHM* is the full-width half maximum of a resonant peak. The second term is also known as the loaded quality factor ($Q_{loaded}$). When the microring resonator is critically coupled to a bus waveguide, $Q_{int} = 2 \times Q_{loaded}$. Critical coupling requires a sufficiently small gap between an optical waveguide and the ring resonator. With i-line lithography, it is challenging to realize gaps smaller than 0.7 µm. Race-track resonators are preferred when the lithography resolution is a limiting factor for small gaps and instead coupling length is increased to reach critical coupling. However, they have abrupt mode changes at the connection interfaces of the straight and bent waveguides, which results in higher losses. Elliptical microring resonators are a better alternative to race-track resonators having a gradual and non-abrupt mode size change in addition to exhibiting enhanced mode coupling due to their elliptical shape. Therefore, we fabricated elliptical microring resonators to estimate the propagation loss of the hybrid waveguide platform. The fabricated elliptical microring resonator has a major radius of 300 µm and a minor radius of 200 µm (an effective radius of $R = 250$ µm [16]), and an *FSR* = 0.95 nm. First, we used a supercontinuum light source (NKT SuperK EXTREME, EXR-4) and an optical spectrum analyzer (Yokogawa, AQ6370B) to measure the FSR and resonance wavelengths of the fabricated resonators (Fig. 4a). To measure the full-width-at-half-maximum (FWHM) of the resonance peaks we used a custom-developed tunable laser (TE polarized, $\lambda_c = 1500$ nm, bandwidth=2.1 nm, MHz linewidth) combined with a battery-operated photodiode (DET08CFC, Thorlabs). The FWHM value of the resonance peak at $\lambda = 1548.68$ nm was measured as 4.5 pm, which corresponds to $Q_{int} = 4.1 \times 10^5$ (Fig. 4b). Inserting the corresponding $Q_{int}$ and the effective radius of the elliptical microring to Equation (1), the optical loss in the ring resonator was calculated to be 0.7 dB/cm. Note that this loss value indicates the upper limit of the loss, which includes both the propagation loss and self-coupling loss of the microring resonator since spectral linewidth is affected by both, and disentangling their effects is not very straightforward. The current loss value is mainly dominated by the material absorption of the SU8 layer. The durability of the polymer layer was checked by measuring the *Q*-factor of the elliptical ring resonators 6 and 12 months after the fabrication and no significant difference was observed.

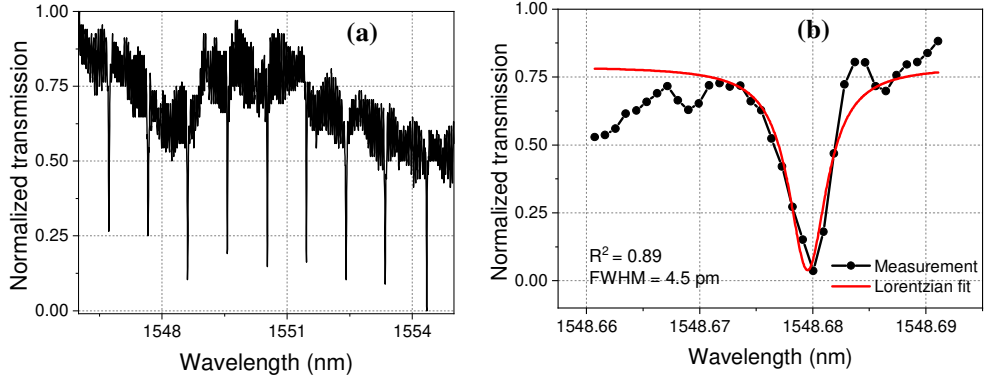

**Figure 4. (a)** Transmission measurement results of the fabricated elliptical ring resonator using a broadband light source and OSA. **(b)** The measurement of the resonance peak at 1548.68 nm using a tunable laser and photodetector. Redline is the Lorentzian fit of the experimental data.

### 3. Design and experimental results of PIC components

*3.1 Arrayed waveguide grating (AWG)*

One of the PIC components that we tested in our hybrid waveguide platform is AWG, which is a planar dispersive device that is generally used as an on-chip spectrometer in various applications [3]. As a proof-of-principle, we designed and fabricated an 8-channel AWG device centered at $\lambda_c$ = 1550 nm with a channel spacing of $\Delta\lambda$ = 1.2 nm and a bandwidth of FSR = 8.4 nm. The remaining design parameters of the devices were calculated using the standard equations for AWGs [18] as follows; diffraction grating ($m$) = 161, focal length ($R_O$) = 310 µm, path length increment ($\Delta L$) = 127 µm, and the number of arrayed waveguides ($M$) = 24.

The AWG layout shown in Fig. 5a was fabricated using our hybrid waveguide platform. The red light propagation in the AWG device is shown in Fig. 5b. The transmission spectra of the AWG output waveguides were measured using the broadband light source and the OSA and normalized to a straight waveguide transmission (Fig. 5c). As predicted, each channel works as a band-pass wavelength filter. The measured values of resolution and FSR are consistent with the calculations. The excess loss is around 2.4 dB, which is mainly due to fabrication limitations (i.e. the large gap between adjacent arrayed waveguides). The adjacent crosstalk value varies between 3 dB (left of the spectrum) and 7 dB (middle of the spectrum). The overall crosstalk was around 13 dB.

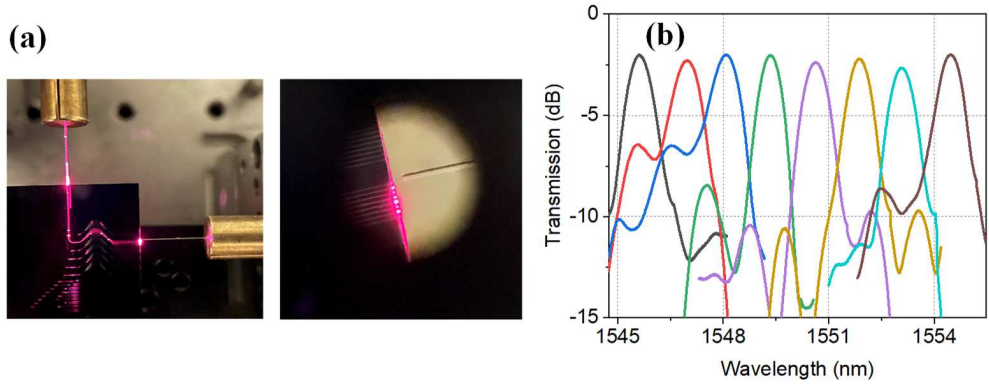

**Figure 5. (a)** Red light propagation in the AWG device (left) and the focused spots at the output channels (right). **(b)** Transmission measurement results of the fabricated AWG device.

*3.2 Ultra-broadband optical splitter*

One of the novel PIC components that were developed using this hybrid technology is an ultra-broadband 3-dB power splitter. Optical power splitters are indispensable components of various PIC concepts [19-21]. Several approaches have been investigated to realize wavelength-insensitive 3-dB couplers; however, they mainly provide a limited bandwidth or suffer from a large footprint and excess loss [22-25]. Therefore, there is a demand for compact and ultra-broadband 3-dB power splitters that can be used both as a stand-alone device and also as the core building block of an optical switch. Recently, we developed a new type of an on-chip 2×2 optical power splitter that works over the 1300-1600 nm wavelength range with a 3-dB splitting ratio [26]. We validated the working principle of this coupler using the standard $Si_3N_4$ technology of Lionix International and obtained a wavelength-flattened response. Below we will present the specific design parameters of this coupler in our hybrid waveguide platform and a more detailed coupler design can be found in Ref. [26].

The coupler is an asymmetric non-uniform directional coupler that consists of two tapered waveguides. One of the coupler arms is shifted by a given amount (i.e. $L_{shift}$) in the propagation direction (Fig. 6b), which results in a larger wavelength-insensitive 3dB-response compared to a standard (not-shifted) coupler (Fig. 6a). Some design parameters were preset considering the fabrication limitations (gap, $g$), single-mode operation, and compactness criteria. First, the length of the tapered section was set to be $L_{taper}$ = 500 µm, which is sufficiently long for a semi-adiabatic transition. The widths of the input and output waveguides of the tapered sections were chosen as $w_1 = 1.4$ μm, $w_2 = 1.2$ μm, and $w_3 = (w_1 + w_2)/2 = 1.3$ μm, to satisfy single mode operation while minimizing sensitivity to fabrication. The separation between waveguides was chosen as $g = 1.5$ μm due to fabrication limitations. To find the optimum $L_{shift}$ value, the transmission response of the non-uniform coupler was simulated in the range of 0-150 μm with 5-μm steps. The absolute deviation of the splitting ratio from the 3-dB value was extracted for each $L_{shift}$ value. For brevity, only the results corresponding to six different values are depicted in Fig. 6c. The optimum shift value was extracted as $L_{shift} = 100$ μm. To compare the improvement provided by this shift, two asymmetric non-uniform couplers were designed and fabricated; one with $L_{shift} = 0$ (not shifted coupler) and another one with $L_{shift} = 100$ μm (shifted coupler). The overall device size is 800 μm including in/out S-bend waveguides. The simulation results of the couplers with $L_{shift} = 0$ and $L_{shift} = 100$ μm are given in Fig. 6d. According to these results, the optimum shift value of $L_{shift} = 100$ μm improved the wavelength flatness of the coupler significantly without increasing the device size. The fabricated devices were characterized using the broadband light source and the OSA. The excess loss was obtained as 0.5 dB by dividing the spectrum of the coupler by the spectrum of a straight waveguide with a similar optical length. The 3dB bandwidth of 300 nm was measured as given in Fig. 6e.

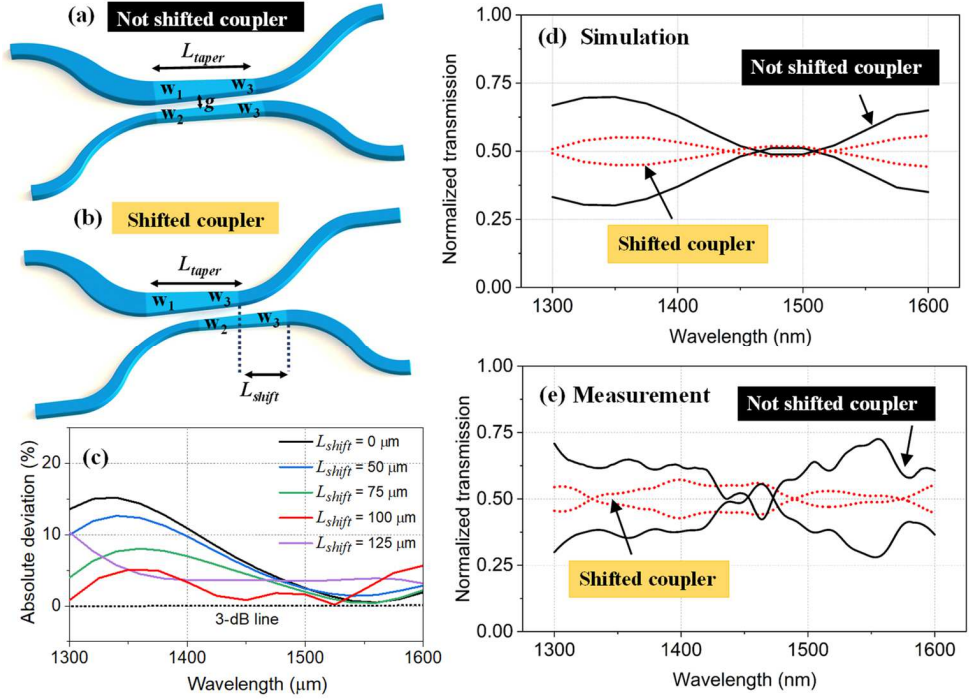

**Figure 6.** Ultra-broadband 3-dB directional coupler. Schematic of the **(a)** not shifted coupler, i.e. $L_{shift} = 0$ **(b)** the shifted coupler $L_{shift} > 0$. **(c)** Absolute deviation from 50/50 coupling ratio for different $L_{shift}$ values. Note that for $L_{shift} = 100$ μm, the deviation gets smaller. **(d)** Simulation and **(e)** measurement results of the not shifted ($L_{shift} = 0$) coupler and shifted ($L_{shift} = 100$ μm) coupler.

### 3.3 Coupled microcavity resonators

Using this hybrid waveguide technology, we also fabricated two designs of coupled microcavity resonators that show Fano shape resonances in their spectra. The Fano resonance is recognized as an important concept across multiple fields of physics [27]. In photonics, it results from the destructive interference between a narrow discrete mode of the cavity and a broad continuum of background modes. The sharp asymmetric Fano lineshapes find interesting applications in the design of sensors [28], optical switches [29], nonlinear devices [30], and lasers [31]. In compound microcavity resonators, it is easier to control the shape and frequency of the Fano resonances dynamically by tuning amplitude and phase differences between cavities [32-34]. Moreover, compound cavities have great potential in optical biosensing by creating an internal reference point to eliminate environmental effects [35]. In this work, we experimentally demonstrated two designs of coupled microcavity resonators. One compound resonator is formed by coupling a microring resonator with a Fabry-Pérot (FP) cavity resonator (FP-ring) whereas the other one is formed by coupling two microring resonators via a Y splitter (Y-ring).

### 3.1.1 FP-ring

In the FP-ring structure, the two-mirror FP cavity is formed by combining two Sagnac loop mirrors with a pulley-type microring resonator (Fig. 7a). Sagnac loop mirrors have broadband *and* accurately-controlled reflectivity, which makes it easier to create Fano resonances over a broad wavelength range. We performed variable finite-difference time-domain (varFDTD) simulations (Lumerical Inc.) to predict the behavior of the compound microresonators. The transmission spectrum of the FP-ring device exhibits asymmetric Fano-like non-Lorentzian resonances that are superimposed on a background defined by the FP oscillations. Fano-like resonances can be observed when the resonant frequency of the microring does not coincide with a maximum of the FP resonances.

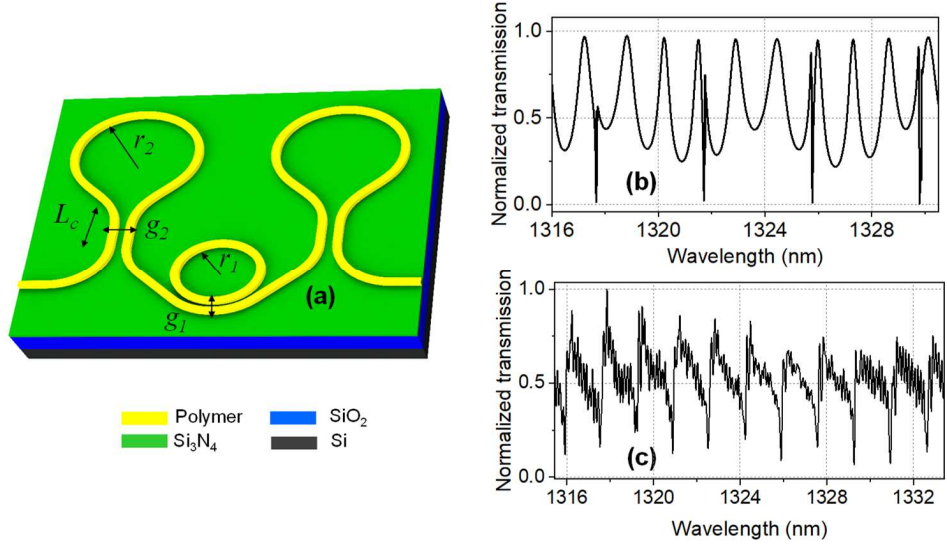

**Figure 7. (a)** Schematic FP-ring structure. **(b)** The FDTD simulation results and **(c)** the transmission measurement results. Due to memory limitations, the radii of the microring resonators were chosen smaller than the fabricated devices (i.e. 50 µm). Except for the FSR difference, the spectra represent the behavior of the device very well.

The FP-ring structure can be analyzed using the transfer matrix method by combining the transfer functions of each component as described in Ref. [33]. In contrast to previous work, here we utilized Sagnac loop mirrors instead of end facet reflections of the chip to form the FP cavity, which makes the design more flexible. The amplitude reflection coefficient $r_S$ of the Sagnac loop mirror can be determined by using the superposition of fields as:

$$r_S = 2j\sqrt{K(1-K)}\exp[(-\gamma + j\beta)L_1], \qquad (3)$$

where $\beta$ is the propagation constant of the waveguide, $\gamma$ is the loss, $L_l$ is the optical path length of the loop waveguide (radius $r_2$) and $K$ is the coupling ratio, which is defined by the coupling constant, $\kappa$, and the coupling length, $L_c$, of the directional coupler. For $K = 0.5$, i.e. 3-dB coupler, the Sagnac loop mirror becomes a perfect reflector as expected. Figure 7a shows the schematic of the FP-ring structure, which consists of a pulley-type microring resonator (radius, $r_1$), and two Sagnac loop mirrors comprised of symmetric directional couplers (length $L_c$ and gap $g$) and circular waveguide sections (radius, $r_2$). The reflectivity of the Sagnac loop mirror is mainly defined by the directional coupler parameters as given in Equation (3). We chose the reflectivity of the loop mirror as $R = 0.6$ by presetting $L_c = 200$ μm and $g_2 = 1.5$ μm, and optimizing them through varFDTD simulations. The radius of the microring resonator was chosen to be $r_1 = 100$ μm. However, due to memory issues, we had to simulate a smaller radius value of $r_1 = 50$ μm, which only increased the FSR but provided a similar output spectrum to the experiments (Fig. 7b). The FSR of the microring resonator and the FP cavity was calculated to be $FSR_{Ring} = 1.8$ nm and $FSR_{FP} = 0.13$ nm, respectively at the central wavelength of 1320 nm. Due to limitations on the fabrication resolution, $g_1$ was chosen to be 1.5 μm. The measurement results are given in Fig. 7c. As it was predicted by the simulations and the theory, we observed Fano resonances in the transmission spectrum. The FSR of the FP and microring resonator was measured as $FSR_{Ring} = 1.8$ nm and $FSR_{FP} = 0.12$ nm, respectively.

### 3.1.2 Y-ring

The schematic of the Y-ring structure is given in Fig. 8a. It consists of two microring resonators of the same radius coupled via a modified blunt Y splitter. The transmission spectrum shows Fano-like resonances centered around the 1300 nm wavelength range. These Fano-like resonances originate from the coupling between the microring resonator and the bent waveguide (providing feedback and thereby creating a second cavity). The theoretical analysis of this specific configuration was given in Ref. [36], where a U-shaped feedback waveguide was utilized to form the second cavity.

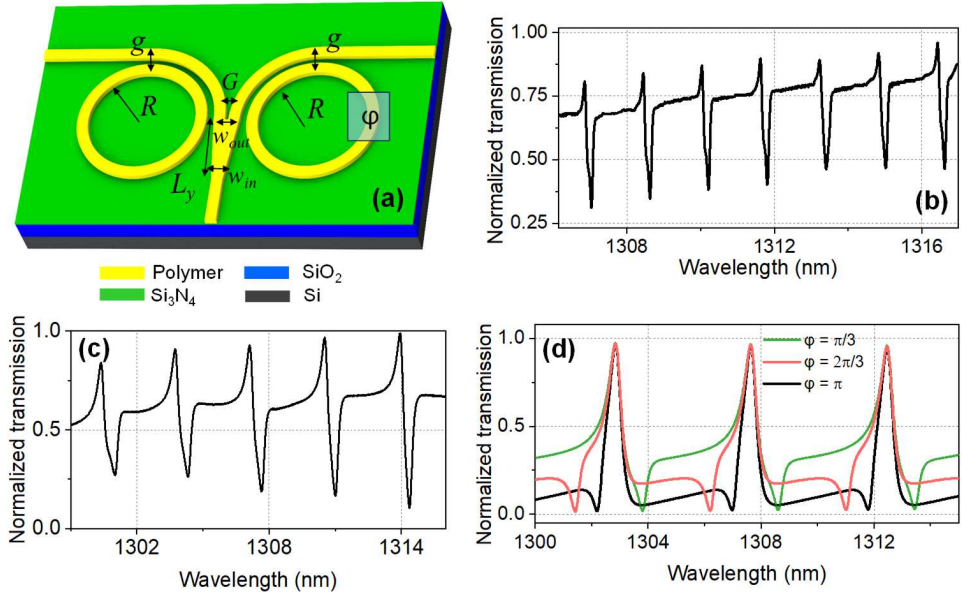

**Figure 8. (a)** Schematic of the Y-ring structure: double microring resonators coupled via a modified blunt Y coupler, **(b)** its transmission measurement results, and **(c)** its FDTD simulation results. **(d)** The shape change of the Fano resonances when the phase of one of the microring resonators is varied. Due to memory limitations, the sizes of the microring resonators were chosen smaller than the fabricated devices (i.e. 60 μm). Except for the FSR difference, the spectra represent the behavior of the device very well.

For a given feedback waveguide length, the shape of the transmission peaks is mainly defined by the gap between the bent waveguide and the ring resonator. For gaps $g \geq 1.5$ µm, asymmetric Fano-like line shapes were obtained (Fig. 8c). We set the length of the bending waveguide as 314 µm, the radius of the microring resonator as $R = 100$ µm, and the gap as 1.5 µm, which resulted in Fano-like resonances. In our configuration, the dual microring resonators are coupled to each other via a modified blunt Y splitter [37]. Compared to conventional Y splitters, they exhibit lower loss and lower susceptibility to fabrication variations. The input and output waveguide widths of the Y splitter are $w_{in} = 1.5$ µm and $w_{out} = 5$ µm, respectively and the gap between output waveguides is $G = 1.5$ µm. The length of the adiabatic region is $L_y = 200$ µm which ensures the lossless transition between input and output waveguide modes. The behavior of the Y-ring structure was simulated using the varFDTD method. A smaller radius, i.e. $R = 60$ µm was used in the simulations due to memory limitations. The results are given in Fig. 8c. As expected, Fano resonances were observed in the transmission response. Using our hybrid waveguide platform, the Y-ring design was fabricated and characterized. The experimental results (spectral shape) match perfectly with the simulation results (Fig. 8b). The FSR of the coupled resonator was obtained as 1.7 nm. We also theoretically analyzed the effect of phase delay between microring resonators on the transmission response. The simulation results show that the shape of the Fano resonances dramatically changes when the phase difference between two microrings is varied between $\varphi = \pi/3 - \pi$ as seen in Fig. 8d. Interestingly, the dip of the Fano resonances stays constant while the peak part is shifting significantly, which makes it a very good candidate for self-referenced optical sensing [35].

## 4. Conclusions

In conclusion, we devised a new approach based on combining weakly-confined optical modes to create highly-confined optical waveguides. A potentially slightly lossy, low refractive index loading medium is combined with a guiding layer of higher refractive index. The thicknesses are chosen such that the layer systems associated with the guiding slab, for absent load, and with the load only, both support weakly confined (slab)-modes close to the respective cutoff. Here the choice of layer thickness turns out to be of critical importance. We tested our approach on a material combination of a SU8 polymer and a $Si_3N_4$ guiding layer. Compact and low-loss waveguides, with a minimum bend radius of 90 µm, for acceptable/negligible bend losses of 0.005 dB/90º, and optical losses of 0.7dB/cm have been realized. We expect that the attenuation can be further reduced by using other low-loss polymers. Measurements on integrated optical micro-ring filters, an arrayed waveguide grating, and a broadband 3dB power splitter demonstrate the capabilities of this highly promising novel waveguide platform. In particular, the technique also enables fast and inexpensive prototyping of new PIC concepts, without sacrificing the PIC performance. We thus expect this new waveguide platform to make a significant difference in various emerging fields of integrated photonics.

**Funding.** Physics2Market

**Acknowledgments.** The authors thank Prof. Atilla Aydinli for the fruitful discussions.

**Disclosures.** The authors declare no conflicts of interest.

**Data availability.** The data that support the findings of this study are available from the corresponding author upon reasonable request.